\documentclass[10pt]{article}
\usepackage[dvips]{graphicx}
 \usepackage{amsmath}
 \usepackage{amsxtra}
\begin{document}
\title {ON AN  ELEMENTARY  DERIVATION OF THE HAMILTON-JACOBI EQUATION FROM
THE SECOND LAW OF NEWTON.}
\bigskip
\author{Alex Granik\thanks{Department of
Physics, University of the
Pacific,Stockton,CA.95211;~E-mail:~agranik@uop.edu}}
\date{}
\maketitle
\bigskip
\begin{abstract}
It is shown that for a relativistic particle moving in an
electromagnetic field  its equations of motion written in a form
of the second law of Newton can be reduced with the help of
elementary operations to the Hamilton-Jacobi equation. The
derivation is based on a possibility of  transforming the equation
of motion to a completely antisymmetric form. Moreover, by
perturbing the Hamilton-Jacobi equation we obtain the principle of
least action.\

The analogous procedure is easily extended to a general
relativistic motion  of a charged relativistic particle in an
electromagnetic field. It sis also shown that the
special-relativistic Hamilton-Jacobi equation for a free particle
allows one to easily demonstrate the wave-particle duality
inherent to this equation and, in addition, to obtain the
operators of the four-momentum whose eigenvalues are the classical
four-momentum
 03.20.+i, 03.30.+p
\end{abstract}
In analytical mechanics we arrive at Newton's second law (the
experimentally verified phenomenological equations with the
observable parameters) by postulating the principle of least
action. In turn, the action $S$ obeys the Hamilton-Jacobi
equation. The latter is a partial differential equation of the
first order.\

A transition from Newton's second law to the Hamilton-Jacobi
equation can be achieved with the help of the
 algorithm for transforming a system of ordinary
differential equations into a partial differential equation.
Despite the fact that such transformation algorithm is well-known
(e.g., \cite{CH}) the actual transformation of the equations of
motion of a charged relativistic  particle in the electromagnetic
field into a respective PDE (the Hamilton-Jacobi equation) is not
quoted in the physical literature to the best of our knowledge.The
usual approach to the problem of derivation of the Hamilton-Jacobi
equation is to heuristically introduce classical action $S$ and to
vary it (for fixed initial and final times).\\

Here we provide an elementary derivation of the Hamilton-Jacobi
where the concept of action emerges in a natural way by
considering the momentum as a function of both temporal and
spatial coordinates. This can be  seen by considering first a
non-relativistic classical particle moving from p.$A$ to p.$F$
(see Fig.$1$). The particle can do that by taking any possible
paths connecting these two points. Therefore for any fixed moment
of time, say $t=1$ the momentum would depend on the spatial
coordinate, that is $\vec{p}=\vec{p}(\vec{x},t)$. In a sense we
have replaced watching the particle evolution in time by watching
the evolution of its velocity (momentum) in space and time.This
situation is analogous to the Euler's description of motion of a
fluid (an alternative to the Lagrange description). The other way
to look at that is to consider a "flow" of an "elemental" path and
describe its "motion" in terms of  its coordinates and velocity
(determined by a slope of the path at a given point).\\

This allows us to represent Newton's second law for a particle
moving in a conservative field $U(\vec{x})$  as follows
\begin{equation}
\label{1} \frac{d\vec{p}}{dt}=\frac{\partial{\vec{p}}}{\partial
t}+\frac{1}{m}({\vec{p}}\cdot\vec{\nabla})\vec{p}=-\nabla U
\end{equation}
We apply $curl$ to both sides of this equation and get
\begin{equation}
\label{2} curl\frac{d\vec{p}}{dt}=\frac{\partial}{\partial
t}{curl\vec{p}}+\frac{1}{m}curl(\vec{p}\cdot\vec{\nabla})\vec{p}=0
\end{equation}

Using the vector identity
\begin{equation}
\label{a}
(\vec{a}\cdot\vec{\nabla})\vec{a}\equiv\frac{\vec{\nabla} a^2}{2}
+curl\vec{a}\times\vec{a} \end{equation}
 we rewrite Eq.(\ref{2})
\begin{equation}
\label{3} \frac{\partial}{\partial
t}{curl\vec{p}}+\frac{1}{m}curl(curl\vec{p}\times\vec{p})=0
\end{equation}
\begin{figure}
 \begin{center}
 \includegraphics[width=6cm, height=6cm]{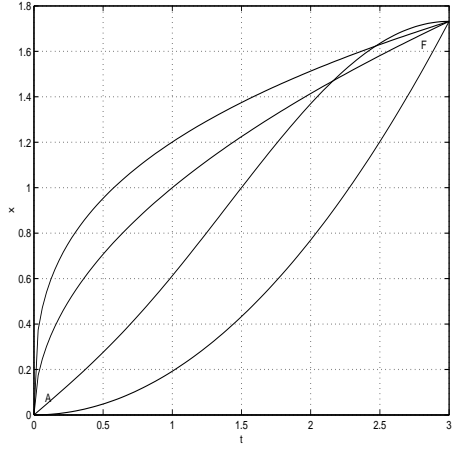}
 \caption{\small A few paths of a path set connecting the initial and the final points
travelled by a particle in $t= 3sec$. It is clearly seen that
particle's  velocity ( momentum) is a function of both coordinate
$x$ and time $t$}
 \end{center}
 \end{figure}

One obvious solution to Eq.(\ref{3}) is $$curl\vec{p}=0$$ similar
to an {\it irrotational} motion in Euler's picture of a fluid
motion.\\

 Eq.(\ref{3}) implies
\begin{equation}
\label{4} \vec{p}=\nabla S
\end{equation}
where $S(\vec{x},t)$ is some scalar function.  Generally speaking,
we can choose the negative value of $\nabla S$. The conventional
choice is connected with the fact that the corresponding value of
the kinetic energy has to be positive.Upon substitution of
Eq.(\ref{4}) into Eq.(\ref{1}) we obtain with the help of
Eq.(\ref{a}) the following equation
\begin{equation}
\label{5} \nabla\{\frac{\partial S}{\partial
t}+\frac{1}{2m}(\nabla S)^2+U\}=0
\end{equation}
In turn Eq.(\ref{5}) means that
\begin{equation}
\label{6} \frac{\partial S}{\partial t}+\frac{1}{2m}(\nabla
S)^2+U=f(t)
\end{equation}
where $f(t)$ is some function of time. By introducing a new
function
$$S'=S-\int f(t)dt$$ we get from (\ref{5}) the Hamilton-Jacobi equation
with respect to the function $S'$ (representing the classical
action):
\begin{equation}
\label{insert} \frac{\partial S'}{\partial t}+\frac{1}{2m}(\nabla
S')^2+U=0
\end{equation}
\\

If we use relation (\ref{4}), $\vec{v}=\vec{p}/m$, and  drop the
prime at $S'$, the Hamilton-Jacobi equation can be rewritten as
follows
\begin{equation}
\label{7} \frac{\partial S}{\partial t}+\vec{v}\cdot\nabla
S=\frac{mv^2}{2}-U
\end{equation}
Since by definition $$\frac{dS}{dt}=\frac{\partial S}{\partial
t}+\vec{v}\cdot\nabla S$$ we obtain from (\ref{7}) the expression
for the action $S$ by integrating (\ref{7}) from p. $A$ to p.$F$
\begin{equation}
\label{7a}
S=\int_{t_A}^{t_F}(\frac{mv^2}{2}-U)dt\equiv\int_{t_A}^{t_F}
L(\vec{x},\vec{v},t)dt
\end{equation}
where $L(\vec{x},\vec{v},t)=mv^2/2-U$ is the lagrangian of a
particle of mass
$m$.\\

Now we can arrive at the principle of least action (without
postulating it {\it a priori}) directly from the Hamilton-Jacobi
equation. To this end we subject the action $S$ to small
perturbations $\delta S \ll S$ and (by dropping the term $(\nabla
\delta S)^2$) get from (\ref{insert}) the equation with respect to
$\delta S$
\begin{equation}
\label{i2} \frac{\partial \delta S}{\partial t}+\frac{1}{m}(\nabla
S)\cdot(\nabla\delta S)=0
\end{equation}
Since $\nabla S/m =\vec{v}$ Eq.(\ref{i2}) represents the
substantial derivative of $\delta S$:
\begin{equation}
\label{i3} \frac{d\delta S}{dt}=0
\end{equation}
This means that
\begin{equation}
\label{i4}
 \delta S= const
\end{equation}
Thus on one hand,  for a specific function $S$ satisfying the
Hamilton-Jacobi equation the respective perturbations $\delta S =
const$. On the other hand, according to Eq. (\ref{7a}) the action
$S$ is defined on a set of all possible paths connecting point $A$
and point $F$. This means that perturbations $\delta S$ correspond
to perturbations of all these path.\\

However, only for one of these paths $\delta S =const$, according
to (\ref{i4}). To determine this constant we take into account
that at the fixed points $A$ and $F$ the paths are also fixed,
that is the respective perturbations $\delta S = 0$ at these
points. Therefore only for the specific path determined by the
Hamilton-Jacobi equation ( that is by the second law of Newton)
$\delta S =0$, thus yielding the principle of least action:
\begin{equation}
\label{i5} \delta\int_{t_A}^{t_F}L(\vec{x},\vec{v},t)=0
\end{equation}\\

The above derivation serves as a guide for a derivation of the
Hamilton-Jacobi equation for a relativistic particle of charge $q$
and mass $m$ moving in the electromagnetic field. Our approach is
to reduce the respective equations of motion to the form which
would be analogous to an irrotational motion in Euler's picture.
The very structure of the space-time metric allows one to arrive
at the required result in a natural way.\\

Therefore we begin with the second law of Newton for a
relativistic charged particle of a charge $q$ and mass  $m$ moving
in the electromagnetic field:
\begin{equation}
\label{8} \frac{dp^{\alpha}}{dt}=  q [ E^\alpha +
\epsilon^{\alpha\beta\gamma}v^{\beta}B^{\gamma}]
\end{equation}

where Greek  indices $\alpha,\beta,\gamma, . . .$ take the values
$1, 2, 3$, $\epsilon^{\alpha\beta\gamma}$ is the absolutely
antisymmetric tensor of the third rank, $p^{\alpha}=
mv^{\alpha}/(1-v^{\delta}v^{\delta})^{1/2}$ is the momentum of the
particle, $E^{\alpha}$ is the electric field ,
$v^{\alpha}=\vec{v}$ is the velocity of the particle and
$B^{\alpha}$ is
the magnetic field.\\

For the subsequent analysis we cast Eq.(\ref{8}) into the standard
co - and contra-variant forms.  To this end we use the metric
$g^{ik}=g_{ik}=[1,-1,-1,-1]$ and use units where the speed of
light is $c=1$. In this metric $x^0 = x_0 = t$,
$x^{\alpha}=\vec{x}=-x_{\alpha}$, the four- potential
$A^i(A^0,A^{\alpha})$ whose scalar part $A^0 = \phi$ (where $\phi$
is the scalar potential) and $A^{\alpha}\equiv\vec{A}$ is the
vector potential, and the roman indices $i, j, k$ ,. . . take the
values $0, 1, 2, 3$,. From the Maxwell equations then follows
(e.g.\cite{LL}) that the electric field $E^{\alpha}$ intensity and
the magnetic induction $B^{\alpha}$ are
\begin{equation}
\label{9} E^{\alpha} = - (\frac{\partial A^0}{\partial
x^{\alpha}}+\frac{\partial A^{\alpha}}{\partial x^0} )
\end{equation}
\begin{equation}
\label{10} B^{\alpha} = \epsilon^{\alpha\beta\gamma}
\frac{\partial A^{\gamma}}{\partial x^{\beta}}
\end{equation}

Using (\ref{10}) we express the second term on the right-hand side
of  Eq. (\ref{8}) in terms of the vector-potential
$A^{\alpha}\equiv\vec{A}$
\begin{equation}
\label{11} \epsilon^{\alpha\beta\gamma}v^{\beta}B^{\gamma}
=\epsilon^{\alpha\beta\gamma}\epsilon^{\gamma\delta\lambda}\frac{\partial
A^{\lambda}}{\partial x^{\delta}}=v^{\beta}(\frac{\partial
A^{\beta}}{\partial x^{\alpha}}-\frac{\partial
A^{\alpha}}{\partial x^{\beta}} )
\end{equation}

Substitution of  (\ref{10}) and(\ref{11}) into  (\ref{8}) yields
\begin{equation}
\label{12}
 \frac{dp^{\alpha}}{dx^0}=q[-(\frac{\partial
A^0}{\partial x^{\alpha}}+\frac{\partial A^{\alpha}}{\partial
x^0})+\beta^{\gamma}(\frac{\partial A^{\gamma}}{\partial
x^{\alpha}}-\frac{\partial A^{\alpha}}{\partial x^{\gamma}})]
\end{equation}
where  $\beta^{\gamma} = v^{\gamma}$.\\

If we use in (\ref{12}) the antisymmetric tensor $F^{ik}$
$(e.g.,\cite{LL})$
\begin{equation}
\label{13}
F^{ik}=\frac{\partial A^k}{\partial x_i}-\frac{\partial
A^i}{\partial x_k}
\end{equation}
the relation between contra- ($A^{\alpha}$) and co-variant
($A_{\alpha}$) vectors ($A^{\alpha}=-A_{\alpha}$), introduce the
space-time interval
$$ds\equiv dt\sqrt{1-\beta^{\alpha}\beta^{\alpha}}\equiv dt\sqrt{1-\beta^2}$$ and the four-velocity
$$u^i(u^0=1/\sqrt{1-\beta^2},~~~~~~~~ u^{\alpha}=-u_{\alpha}=\beta^{\alpha}/\sqrt{1-\beta^2}),$$
we get
\begin{equation}
\label{14} \frac{dp^{\alpha}}{ds}=qF^{\alpha
k}u_k=-qF^{k\alpha}u_k
\end{equation}

As a next step, we find the zeroth components of Eqs.(\ref{14}).
Using the special-relativistic identity for the momentum
$p_i=mu_i$$$ p_ip^i=m^2$$ we find
\begin{equation}
\label{15}
 p_0\frac{dp^0}{ds}\equiv
 -p_{\alpha}\frac{dp^{\alpha}}{ds}=p^{\alpha}\frac{dp^{\alpha}}{ds}
 \end{equation}

Upon insertion of (\ref{15}) into (\ref{14})we obtain
\begin{equation}
\label{16}
p^{\alpha}\frac{dp^{\alpha}}{ds}=qp^{\alpha}[F^{\alpha\beta}u_{\beta}+F^{\alpha
0}u_0]
\end{equation}
On the other hand, since $F^{ik}=-F^{ki}$( $F^{00}=F_{00}=0$)

$$p^{\alpha}u_{\beta}F^{\alpha\beta}=0$$
Hence from (\ref{15}) and (\ref{16}) follows that
\begin{equation}
\label{17} \frac{dp^0}{ds}=qu^{\alpha}F^{\alpha
0}=qF^{0\alpha}u_{\alpha}=qF^{0i}u_i
\end{equation}
Adding Eqs.(\ref{17}) and (\ref{14}) and using the definition of
$F^{ik}$, Eq.(\ref{13}), we arrive at the equation of motion in
the contra-variant form:
\begin{equation}
\label{18} \frac{dp^i}{ds}=qF^{ik}u_k=q(\frac{\partial
A^k}{\partial x_i}-\frac{\partial A^i}{\partial x_k})u_k
\end{equation}
The respective co-variant form follows from raising and lowering
indices in (\ref{18}):
\begin{equation}
\label{19} \frac{dp_i}{ds}=qF_{ik}u^k=q(\frac{\partial
A_k}{\partial x^i}-\frac{\partial A_i}{\partial x^k})u^k
\end{equation}\\

Now we reduce  these equations  to a form similar to the condition
defining an irrotational flow in fluid mechanics. To this end we
rewrite  (\ref{18}) and (\ref{19}) in the following form
\begin{eqnarray}
\label{20} u_k[\frac{\partial}{\partial
x_k}(mu^i+qA^i)-\frac{\partial}{\partial x_i}(qA^k)]=0\nonumber\\
u^k[\frac{\partial}{\partial
x^k}(mu_i+qA_i)-\frac{\partial}{\partial x^i}(qA_k)]=0
\end{eqnarray}

and add to the third term the identity
$$u_k\frac{\partial u^k}
{\partial x_i}=u^k\frac{\partial u_k}{\partial
x^i}\equiv\frac{1}{2}\frac {\partial}{\partial x_i}(u_ku^k)=0
$$

As a result, we get
\begin{equation}
\label{21} u_k[\frac{\partial}{\partial
x_k}(mu^i+qA^i)-\frac{\partial}{\partial x_i}(mu^k+qA^k)]=0
\end{equation}
or equivalently
\begin{equation}
\label{22} u^k[\frac{\partial}{\partial
x^k}(mu_i+qA_i)-\frac{\partial}{\partial x^i}(mu_k+qA_k)]=0
\end{equation}
The expressions in square brackets represent a four-curl of the
four-vector $mu_i+qA_i$ (or $mu^i+qA^i$). Both equations are
identically satisfied if this four-curl is $0$. Once again, this
can be interpreted as the fact that the respective vector field is
irrotational, that is the four-vector $m\vec{u}+q\vec{A}$ (here we
use  notation  $\vec{a}$ for a four-vector) is the four-gradient
of a scalar function, say $- S$
\begin{equation}
\label{23} mu^i +qA^i = - \frac{\partial S}{\partial x_i}
\end{equation}
\begin{equation}
\label{24} mu_i +qA_i = -\frac{\partial S}{\partial x^i}
\end{equation}\\

This scalar function $S$ ( a "potential function") is the
classical relativistic action, and our choice of the sign is
dictated by the consideration that expressions (\ref{23}) must
become the expressions for the momentum and energy in the
non-relativistic limit.\\

To find the explicit expression for $S$ we integrate Eq.
(\ref{23}) [ or (\ref{24})] and obtain:
\begin{equation}
\label{25} S=-\int_a^b(mu^i + qA^i
)dx_i\equiv-\int_a^b(m+A^iu_i)ds
\end{equation}
where $a$ and $b$ are points on the world line of the particle,
$ds=(dx^idx_i)^{1/2}$, and $u_i=dx_i/ds$. Expression (\ref{25})
coincides ( as it should be) with the conventional definition of
the action (introduced on the basis of considerations not
connected to the second law of Newton). It is interesting to note
that in a conventional approach to the action, the term $A^idx_i$
"cannot be fixed on the basis of general considerations alone"
\cite{LL}. Here however this term is "fixed" by the very nature of the equations of motion.\\

Eqs. (\ref{23}) and (\ref{24}) yield the determining PDE for the
function $S$ (the relativistic Hamilton-Jacobi equation for a
charged particle in the electromagnetic field) if we eliminate
$u_i$ and $u^i$ from this equations with the help of the identity
$u_iu^i=1$:
\begin{equation}
\label{26} (\frac{\partial S}{\partial x_i}+ qA^i)(\frac{\partial
S}{\partial x^i}+ qA_i)=m^2,~~~~~~~i=0,1,2,3
\end{equation}
where we have to retain ( in the classical region) only one sign,
either  plus or minus.\footnote{We would like to point out that a
unified way to describe wave and particle phenomena inherent to
the Hamilton-Jacobi equation ( which was the main motivation of
Hamilton) is conventionally demonstrated by comparing it and the
eikonal equation and by showing that they are identical. On the
other hand, as we show in the Appendix, there exists a simple way
to do that directly from the Hamilton-Jacobi equation without
resorting to the eikonal equation.}\\

The usual way to derive the equations of motion (\ref{18}) [or
(\ref{19})] from the action, Eq.(\ref{25}) is to vary it. Here we
follow the well-known procedure of reducing the integration of the
partial differential equation of the first order  to the
integration of a system of the respective ordinary differential
equations  \cite {CH}. In particular, given the Hamilton-Jacobi
equation (\ref{26}) we derive (\ref{18}). To this end we subject
action $S$ to small perturbations $\delta S$
\begin{equation}
\label{27} S = S_0 + \delta S
\end{equation}
and find the equation governing these perturbations. Here $S_0$
must satisfy the original unperturbed Hamilton-Jacobi equation
(\ref{26}) , and $\delta S\ll S_0$.\\

Upon substitution of (\ref{27}) into (\ref{26}) we get with
accuracy to the first order in $\delta S$
\begin{equation}
\label{28} (\frac{\partial S_0}{\partial x_i}+
qA^i)\frac{\partial}{\partial x^i}\delta S+(\frac{\partial
S}{\partial x^i}+ qA_i)\frac{\partial}{\partial x_i}\delta S=0
\end{equation}
 or equivalently
\begin{equation}
\label{29} (\frac{\partial S}{\partial x^i}+
qA_i)\frac{\partial}{\partial x_i}\delta S=0
\end{equation}

Equation (\ref{29}) is a quasi-linear first-order PDE whose
characteristics are given by the following  equations
\begin{equation}
\label{30} \frac{dx_0}{\partial S_0/\partial x^0+ qA_0}=
\frac{dx^{\alpha}}{\partial S_0/\partial x_{\alpha}+ qA^{\alpha}}
\end{equation}

Here the repeated indices do not represent  summation, and $\alpha
= 1, 2, 3$. It is immediately seen that the characteristics of
linearized Hamilton-Jacobi equation (\ref{30}) are the four-
velocity $u^i$:
 \begin{equation}
 \label{33}
 u^i=\frac{1}{m}(\frac{\partial S_0}{\partial x_i}+ qA^i)
 \end{equation}\\

 Inversely, these characteristics are the solutions of the
equations of motion written in a form of the second law of Newton.
To demonstrate that we divide  both sides of (\ref{33}) by $ds$,
use Eqs. (\ref{23}),(\ref{24})  and the  fact that
$d/ds=u_k\partial /\partial x_k$ and obtain
\begin{eqnarray}
\label{34}
 mc\frac{du^i}{ds}=\frac{1}{m}(\frac{\partial S_0}
 {\partial x^k}+ qA_k)\frac{\partial}{\partial x_k}(\frac{\partial S_0}
 {\partial x_i}+ qA^i)\equiv\nonumber\\
 \frac{1}{m}(\frac{\partial S_0}
 {\partial x^k}+ qA_k)[\frac{\partial}{\partial x_k}(\frac{\partial S_0}
 {\partial x_i}+ qA^i)+ q\frac{\partial A^k}{\partial x_i}- q\frac{\partial A^k}{\partial
 x_i}]\equiv\nonumber\\
 \frac{1}{m}(\frac{\partial S_0}{\partial x^k}+
 qA_k)[\frac{\partial}{\partial x_i}(\frac{\partial S_0}
 {\partial x_k}+ qA^k)- q(\frac{\partial A^k}{\partial x_i}-\frac{\partial A^i}
 {\partial x_k})]=\nonumber\\\frac{1}{2m}\frac{\partial}{\partial
 x_i}(u_ku^k)+\frac{1}{m}qu_k(\frac{\partial A^k}{\partial x_i}-\frac{\partial A^i}
 {\partial x_k})=qu_kF^{ik}
  \end{eqnarray}
  that is the second law of Newton, Eq.(\ref{18})\\

Now we  return to the linearized equation (\ref{29}) which we
rewrite in the identical form
\begin{equation}
\label{35} mu_i\frac{\partial}{\partial x_i}\delta S\equiv
\frac{d}{ds}\delta S = 0
\end{equation}
This means that $\delta S = const$ along a certain world line,
singled out  of a continuous set of possible world lines according
to this condition. Without any loss of generality we can take the
above $const=0$. \\

Thus on one hand,  for a specific function $S$ satisfying the
Hamilton-Jacobi equation the respective perturbations $\delta S =
const$. On the other hand, according to Eq. (\ref{25}) the action
$S$ is defined on a set of all possible world lines connecting
world points  $a$ and $b$. This means that perturbations $\delta
S$ correspond to perturbations of all these world lines. However,
only for one of these world lines $\delta S =const$, according to
(\ref{35}). To determine this constant we take into account that
at the fixed world points $a$ and $b$ the world lines are also
fixed, that is the respective perturbations $\delta S = 0$ at
these points. If we apply condition Eq.(\ref{35}) to the action
$S$, Eq. (\ref{23}), the former would $"choose"$ out of all
possible world lines the only one satisfying that condition, that
is we arrive at the classical principle of least action.
\begin{equation}
\label{36}
 \delta\int_a^b(mu^i+ qA^i)dx_i=0
\end{equation}\\

Finally, we demonstrate in an elementary fashion how the same
technique of transforming the equations of motion in the Newtonian
form to the Hamilton-Jacobi equation can be applied to a motion of
a charged particle in general relativity. The equations of motion
of a charged particle in gravitational and electromagnetic field
are \cite{LLRuss}.
\begin{equation}
\label{37} M(u^l\frac{\partial u^i}{\partial
x^l}+\Gamma_{kl}^iu^ku^l)=qg^{im}F_{mk}u^k
\end{equation}
where $$\Gamma_{kl}^i=\frac{1}{2}g^{im}(\frac{\partial
g_{mk}}{\partial x^l}+ \frac{\partial g_{ml}}{\partial
x^k}-\frac{\partial g_{kl}}{\partial x^m})$$ is the Ricci tensor.
The expression $\Gamma_{kl}^iu^ku^l$ is significantly simplified
according to the following identities:\\

$$\Gamma_{kl}^iu^ku^l\equiv u^l\frac{1}{2}[u^k(\frac{\partial
g^{im}g_{mk}}{\partial x^l}-g_{mk}\frac{\partial g^{im}}{\partial
x^l})-u^kg_{lm}\frac{\partial g^{im}}{\partial x^k}
-u^kg^{im}\frac{\partial g^{kl}}{\partial x^m}]\equiv$$

$$-\frac{1}{2}[u^lu_m\frac{\partial g^{im}}{\partial
x^l}+u^lu^k(g_{lm}\frac{\partial g^{im}}{\partial
x^k}+\frac{\partial g_{kl}}{\partial
x_i})]\equiv-\frac{1}{2}[2u^lu_m\frac{\partial g^{im}}{\partial
x^l}+u^lu^k\frac{\partial g_{kl}}{\partial x_i})]\equiv$$

$$-u^l(\frac{\partial g^{im}u_m}{\partial x^l}-g^{im}\frac{\partial
u_m}{\partial x^l})-\frac{1}{2}u^lu^k\frac{\partial
g_{kl}}{\partial x_i}\equiv-u^l\frac{\partial u_i}{\partial
x^l}+u^lg^{ik}\frac{\partial u_k}{\partial x^l}-$$

$$\frac{1}{2}u^l(\frac{\partial g_{kl}u^k}{\partial
x^i}-g_{kl}\frac{\partial u^k}{\partial x^i})\equiv
-u^l\frac{\partial u_i}{\partial x^l}+u^lg^{ik}\frac{\partial
u_k}{\partial x^l}- \frac{1}{2}(u^l\frac{\partial u_l}{\partial
x^i}-u_l\frac{\partial u^l}{\partial x_i})\equiv$$

$$-u^l\frac{\partial u_i}{\partial x^l}+u^lg^{ik}\frac{\partial
u_k}{\partial x^l}- \frac{1}{2}(u^l\frac{\partial u_l}{\partial
x_i}-\frac{\partial u^lu_l}{\partial x_i}+u^l\frac{\partial
u_l}{\partial x_i})\equiv$$

$$-u^l\frac{\partial u_i}{\partial x^l}+u^lg^{ik}\frac{\partial
u_k}{\partial x^l}-u^l\frac{\partial u_l}{\partial
x_i}\equiv-u^l\frac{\partial u_i}{\partial
x^l}+u^lg^{ik}(\frac{\partial u_k}{\partial x^l}-\frac{\partial
u_l}{\partial x^k})$$

If we substitute this result into (\ref{37}) and use the
expression (\ref{13}) for $F^{ik}$, we obtain
\begin{equation}
\label{39} g^{ik}u^l[\frac{\partial }{\partial
x^l}(Mu_k+qA_k)-\frac{\partial }{\partial x^k}(Mu_l+qA_l)]=0
\end{equation}

Equation (\ref{39}) is identically satisfied if  we set
\begin{equation}
\label{40} Mu_k+qA_k=-\frac{\partial S}{\partial x^k}
\end{equation}
where $S$ is the action and we use the negative sign, representing
a conventional choice of positive energies in classical mechanics.
Raising and lowering the indices in (\ref{40}), expressing the
respective $4$-velocities $u_k$ and $u^k$ in terms of
$\partial{S}/\partial{x^k}$, and using the identity
$g^{ik}u_iu_k=1$, we arrive at the Hamilton-Jacobi equation:
\begin{equation}
\label{41} g^{ik}(\frac{\partial S}{\partial
x^i}+A_i)(\frac{\partial S}{\partial x^k}+A_k)=m^2
\end{equation}\\

We have shown that with the help of elementary operations one can
arrive at the Hamilton-Jacobi equation from the phenomenological
second law of Newton, without using of  {\it a priori} defined
action $S$. The latter arises in a natural way as a consequence of
the existence of the "irrotational" solutions to the second law of
Newton.The procedure follows from the fact that for forces
determined by the potential energy $U$, the second law of Newton
has a symmetry which allows us to reduce it to an antisymmetric
form analogous to the form observed in the potential flow of an
ideal fluid. This form lends itself to the introduction of a
certain potential function (action), whose gradient represent a
generalized momentum. Upon introduction of the action back into
the second law of Newton we arrive at the Hamilton-Jacobi
equation. We also show in an elementary fashion the wave-particle
duality inherent in the Hamilton-Jacobi equation. As a result of
this derivation we obtain the energy-momentum operators of quantum
mechanics.
\section{Appendix}
Let us consider a motion of a free relativistic particle of a mass
$m$, whose Hamilton-Jacobi equation (\ref{23})
$$\frac{\partial S}{\partial x^i}\frac{\partial S}{\partial x_i}= m^2$$
has a solution:
\begin{equation}
\label{A1} S=-p_ix^i
\end{equation}
corresponding to the energy-momentum  relation $p_ip^i=m^2$ of the
special relativity.\\

On the other hand, if we introduce the  function \footnote{which
was done for the first time  by E.Schroedinger in his historical
paper \cite{ES} on non-relativistic quantum
mechanics}
\begin{equation}
\label{Aa} S=hLn\Psi
\end{equation}
 ( where $h$ is some constant having the dimension of the
action $S$) the Hamilton-Jacobi equation (\ref{A1}) yields:
\begin{equation}
\label{A2} \frac{\partial\Psi}{\partial
x^i}\frac{\partial\Psi}{\partial x_i}=m^2\Psi^2
\end{equation}
This equation admits the wave solution
\begin{equation}
\label{A3} \Psi=e^{-ik_jx^j}
\end{equation}
where $k^i(\omega, \vec{k})$ is the wave four-vector. The
respective dispersion relation is
\begin{equation}
\label{A4} k_jk^j=m^2
\end{equation}
Thus, on one hand the Hamilton-Jacobi equation describes a free
particle with the momentum-energy four vector $p^i(E,\vec{p})$
and, on the other hand, the same equation describes a
monochromatic wave with the wave four-vector $k^i$\\

Returning back to function $S$ ( Eq.\ref{A2}) we obtain
from(\ref{A3})
\begin{equation}
\label{A5} S=-hik_jx^j
\end{equation}
Comparing Eqs.(\ref{A1}) and (\ref{A5}) we obtain the well-known
relation between the wave four-vector and the four-momentum
vector:
\begin{equation}
\label{A6} k_j=\frac{p_j}{ih}
\end{equation}

Moreover,  relation (\ref{A6}) allows us to reinterpret the
four-momentum vector in terms of the eigenvalues of the certain
differential operators. In fact, from (\ref{A3}) and (\ref{A6})
follows
\begin{equation}
\label{A7} hi\frac{\partial\Psi}{\partial x^j}=p_j\Psi
\end{equation}
which means that we arrive at the relations describing
quantum-mechanical operators of momentum-energy.
\begin{equation}
\label{A8}p_j\rightarrow hi\frac{\partial}{\partial x^j}
\end{equation}


\begin{thebibliography}{99}
\bibitem{CH} R.Courant and D. Hilbert, Methods of Mathematical Physics
(Interscience Publishers, New York,1953), vol.2, pp.106-109
\bibitem{LL} L.Landau and E.Lifshitz, The Classical Theory of Fields
(Pergamon Press, New York, 1971), p.60-61
\bibitem{LLRuss} L.Landau and E.Lifshitz, The Classical Theory of
Fields(in Russian)(State Publ.House of Physical and Mathematical
Literature Moscow, 1962), p.302
\bibitem{ES}E.Schroedinger, Annalen der Physik(4), vol.79 (1926)
\end{thebibliography}
\end {document}